\documentclass[10pt, conference]{IEEEtran}
\usepackage{algorithmicx}
\usepackage[ruled,vlined,linesnumbered]{algorithm2e}
\usepackage{cite}
\usepackage{hhline}
\usepackage{amsmath,mathtools}
\usepackage{amsfonts,amssymb}
\usepackage{mathrsfs}
\usepackage{caption}% http://ctan.org/pkg/caption
\usepackage{multirow}
\usepackage{graphicx} 
\usepackage{multirow}
\usepackage{geometry}
\usepackage{enumitem,color}
\usepackage{algpseudocode}
\begin{document}
\title{\huge \emph{LeFi}: Learn to Incentivize Federated Learning in Automotive Edge Computing \vspace{-0.13in}}

\author{\IEEEauthorblockN{Ming Zhao,Yuru Zhang, Qiang Liu \vspace{-0.16in}}\\
\IEEEauthorblockA{
University of Nebraska-Lincoln\\
qiang.liu@unl.edu}\vspace{-0.45in}
\and
\IEEEauthorblockN{Tao Han \vspace{-0.16in}}\\
\IEEEauthorblockA{
New Jersey Institute of Technology\\
 tao.han@njit.edu}\vspace{-0.45in}
}

\maketitle

\begin{abstract}
Federated learning (FL) is the promising privacy-preserve approach to continually update the central machine learning (ML) model (e.g., object detectors in edge servers) by aggregating the gradients obtained from local observation data in distributed connected and automated vehicles (CAVs).
The incentive mechanism is to incentivize individual selfish CAVs to participate in FL towards the improvement of overall model accuracy.
It is, however, challenging to design the incentive mechanism, due to the complex correlation between the overall model accuracy and unknown incentive sensitivity of CAVs, especially under the non-independent and identically distributed (Non-IID) data of individual CAVs.
In this paper, we propose a new learn-to-incentivize algorithm to adaptively allocate rewards to individual CAVs under unknown sensitivity functions. 
First, we gradually learn the unknown sensitivity function of individual CAVs with accumulative observations, by using compute-efficient Gaussian process regression (GPR).
Second, we iteratively update the reward allocation to individual CAVs with new sampled gradients, derived from GPR.
Third, we project the updated reward allocations to comply with the total budget.
We evaluate the performance of extensive simulations, where the simulation parameters are obtained from realistic profiling of the CIFAR-10 dataset and NVIDIA RTX 3080 GPU.
The results show that our proposed algorithm substantially outperforms existing solutions, in terms of accuracy, scalability, and adaptability.
\end{abstract}


\section{Introduction}
\label{sec:introduction}

% what is federated learning, benefits (privacy, communication -efficient)s
Federated learning is the promising paradigm to achieve large-scale privacy-preserve distributed machine learning.
To collaboratively train a global model, a generic federated learning system involves massive participants (which compute the model update with their local data) and a central parameter server, which aggregates distributed updates in the global model, such as FedAvg~\cite{mcmahan2017communication}.
Instead of sharing the raw training data, only computed model updates (e.g., gradients) will be transmitted to the central server, which substantially alleviates the data privacy concerns and improves communication efficiency.
Hence, there are ever-increasing research efforts in exploring and exploiting the advantages of federated learning in a wide range of application domains, such as healthcare, finance, and retail~\cite{yang2019federated}.

% incentive mechanism,
The incentive mechanism (i.e., incentivizing massive participants to contribute their data) plays a key role in achieving the goal of federated learning~\cite{zhan2020learning}. 
This is based on the observation that individual participants are generally reluctant to participate in federated learning without monetary incentives.
On the one hand, the local training consumes non-trivial computation resources, which would squeeze the limited computation capacity of participants (usually edge devices).
On the other hand, the transmission of the computed model updates requires non-negligible mobile traffic, which may increase the data usage of participants (usually mobile devices).
As a result, the sensitivity function of participants is different concerning monetary incentives. 
In addition, recent findings show that not all participants are equally important, in terms of improving the overall accuracy of the global model, because of their heterogeneous data quantity and quality~\cite{zhan2021survey}, such as independently and identically distributed (IID). 

% existing works in incentive mechanism, what are their challenges
There are extensive existing works in designing incentive mechanisms in the domain of federated learning~\cite{zhan2020learning, kang2019incentive}.
However, these works generally assume that the information of local data (e.g., quality and quantity) in participants is known by the central server, while this assumption may not always hold.
For example, participants can be selfish, and optimize their contribution (e.g., data and computation) dynamically in each federated learning round.
Recent works focus on using model-free approaches (e.g., deep reinforcement learning~\cite{zhan2021incentive, yang2023bara}) to learn and incentivize participants without the need for prior knowledge.
However, they typically use deep neural networks (DNNs) to parameterize their policies, which tend to be low sample efficiency and can hardly scale under dynamic participants over time.
Therefore, it is imperative to design a sample-efficient and model-free approach to incentivize diverse participants in federated learning.

% in this paper
In this paper, we propose the \emph{LeFi} algorithm to learn and incentivize participants, i.e., connected and automated vehicles (CAVs), in automotive edge computing.
The optimization objective is to maximize the overall accuracy of the global model under the total monetary budget, by optimizing the reward weight of individual participating CAVs.
First, we develop a sample-efficient surrogate model to approximate the unknown sensitivity function of participants, according to the accumulative Server-to-CAV interactions over time.
Second, we design a sample-based gradient descent method to update the reward weight of individual CAVs and project the reward weights under the total budget if applicable.
Third, we use Karush-Kuhn-Tucker (KKT) conditions to derive the data selection in each CAV independently.
Then, the central server will receive the actual data size of local training in each CAV, which will be used to update the surrogate model continuously.
We evaluate the proposed algorithm via extensive simulations, where the simulation parameters are obtained by profiling a Convolutional neural network (CNN)-based model under the CIFAR-10 dataset and Nvidia RTX 3080 GPU.
The results show that our proposed algorithm substantially outperforms existing solutions, in terms of convergence, accuracy, and scalability.

\section{System Model}

% describe the model, e.g., define the server and CAVs, how federated learning is working in this paper, what is iid, what is reward allocation, what is data selection
We consider a generic federated learning system, including a central server and a massive number of CAVs denoted as $\mathcal{N} = \{1, 2, ..., N\}$.
To train the global DNN model, e.g., object detection and segmentation, CAVs perceive surrounding environments with diverse onboard sensors (e.g., LiDAR and cameras) and are responsible for the local training process before the calculated gradients are sent to the central server. 
The central server will aggregate the gradients received from CAVs, e.g., averaging, to update the global DNN model.
The global model will be then disseminated to the participating CAVs, e.g., changed parameters, which creates the next round of federated learning.

\textbf{Non-IID Data.} Recent findings~\cite{zhan2021survey} show that the distribution of training data of individual participants (i.e., CAVs) would generate a non-trivial impact on the overall accuracy of the global model. 
The training data is IID, only if its samples are drawn from the same probability distribution, and are independent of one another statistically.
However, the diversified trajectories of individual CAVs will naturally create non-IID training data.
Specifically, Non-IID data can be characterized by the allocation of a proportion of samples for each label to each party according to a Dirichlet distribution\cite{li2022federated}. 
Without loss of generality, we introduce the variable $\pi \in [0, 1]$ to quantify the degree of Non-IID in the training data. 
A smaller $\pi$ value indicates a higher degree of Non-IID, signifying greater disparity in the proportion of samples for each label across parties. 
In contrast, as $\pi$ approaches 1, it suggests that each party possesses an identical proportion of the samples for each label.

\begin{figure}[!t]
	\centering
	\includegraphics[width=2.5in]{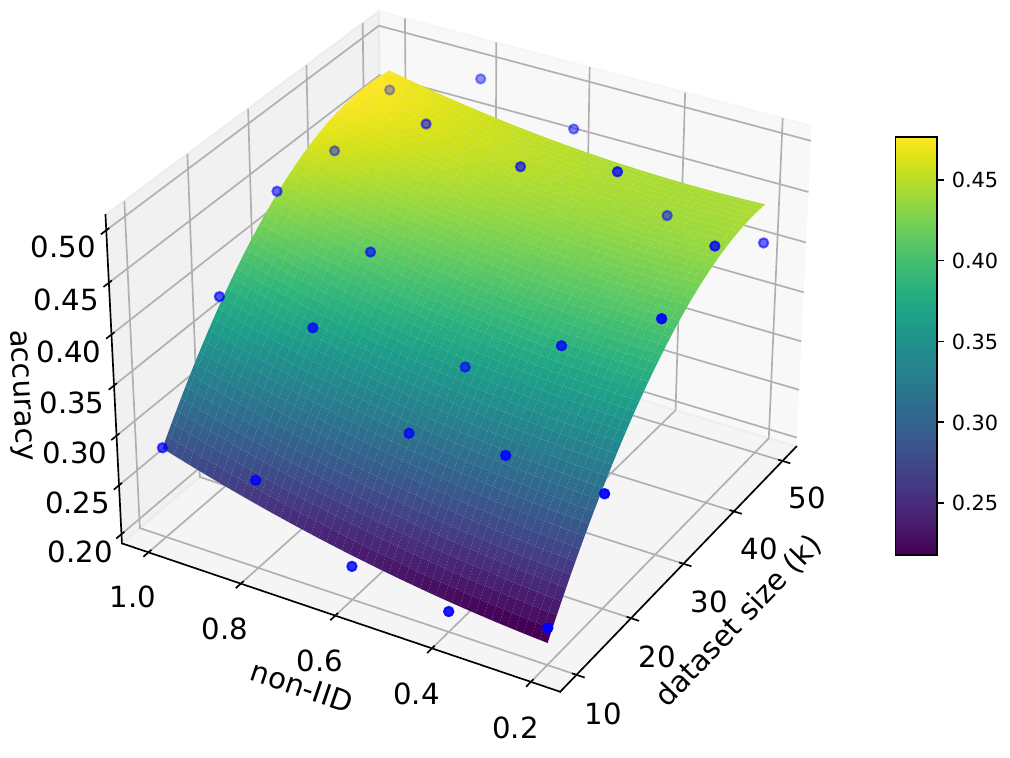}
	\caption{\small Model accuracy varies in different dataset sizes and Non-IID severity}
	\label{fig:acc_noniid_datasize}
\end{figure}

\textbf{Accuracy Model.} 
The overall accuracy of the global model is related to not only the non-IID severity but also the size of training data in individual participating CAVs.
In Fig.~\ref{fig:acc_noniid_datasize}, we show the correlations by training a CNN model over the CIFAR-10 dataset (in total 50K training data) with 5 CAVs (different data sizes and same non-IID severities). 
It can be seen that higher accuracy can be generally achieved by increasing the data size under lower non-IID severity.
Without loss of generality, the accuracy model can be built by fitting these data points with various regression models, such as linear, polynomial and exponential.
However, the non-IID severity of training data in CAVs is mostly different, which generates intractable complexity in regressing the accuracy model, especially under thousands of CAVs, if not more.
For the sake of tractability, we approximately formulate the overall accuracy of the global model as the combination of the individual CAVs' accuracy
\begin{equation}
A_{mean} = \frac{w}{N}\sum\limits_{n \in \mathcal{N}} A(D_n,\pi_n), 
\label{eq:accuracy}
\end{equation}
where $w$ is the weight, which relates to not only the distribution of non-IID severity among all CAVs but also the number of participating CAVs.
Here, $A(D_n,\pi_n)$ is the accuracy function of the $n$-th CAV, which may be regressed with tractable profiling data points.
% \begin{equation}
% \begin{split}
% A_{system}=-0.00004043806774051300D_n^2 + -0.02684348804032874597  \pi_n^2 + 0.00242842147823432676D_n\pi_n + 0.00409433036010190595D_n + 0.02588271138938123828 \pi_n + 0.09225880021658927621
% \end{split}
% \end{equation}
% a_1 D_n^2 + a_2 \pi_n^2 +  a_3 D_n\pi_n + a_4 D_n + a_5 \pi_n+a_6,
% where $a_1 = -4e-5, a_2 = -2.7e-2, a_3 = 2.4e-3, a_4 = 4e-3, a_1 = 2.6e-3, a_1 = 9.2e-2$ with the root mean square error (RMSE) of 0.019.

\textbf{Incentive Mechanism.} 
To improve the overall accuracy of the global model, it is essential to incentivize the contribution of selfish CAVs, in terms of computation resources and unique training data.
As we quantified the aforementioned accuracy model, we designed the incentive mechanism by offering monetary rewards to individual CAVs.
In particular, we define the reward function for the $n$-th CAV as
\begin{equation}
f(D_n) = \alpha_n(1-e^{-\beta_n D_n}),
\label{eq:reward}
\end{equation}
where $\alpha_n$ denotes the reward weight, $D_n$ is the data size of the CAV, and $\beta_n$ is a pre-defined non-negative hyperparameter.
The rationale behind this is that, by optimizing and allocating the reward cap (i.e., $\alpha_n$), the reward function $f(D_n)$ will incentivize the $n$-th CAV to contribute more training data. 
In addition, using the exponential-related functions (rather than linear) can represent a wide range of correlations between the size of training data and the reward. %, such as marginal effects.

\begin{figure}[!t]
	\centering
	\includegraphics[width=2.5in, height=1.7in]{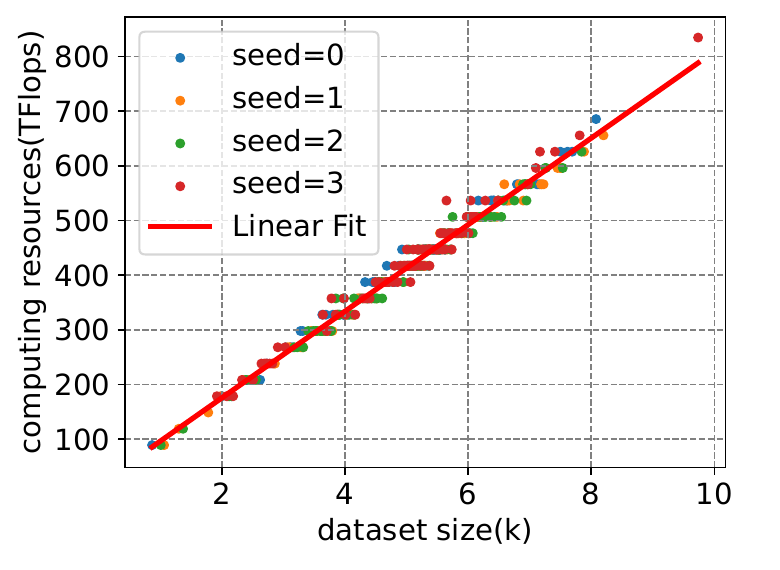}
	\caption{\small Computing resources needs in different dataset sizes }
	\label{fig:timecomplexity}
\end{figure}

\textbf{Selfish Participating CAVs.}
The local training process and sharing of data require non-negligible costs in each CAV, e.g., computing resources and mobile traffic.
Hence, it is reasonable to assume that each CAV is selfish in participating the federated learning, e.g., monetary rewards to compensate for local resource consumption.
In particular, we profile the computational complexity of the local training process under different training data sizes in Fig.~\ref{fig:timecomplexity}, by using the metric of TFLOPS (Tera Floating-point operations per second).
By fitting the profiled data point with linear regression, we obtain the computation complexity $C(D_n) = pD_n + q$, where $p=79.1259$ and $q=17.6219$. % with the RMSE of 13.56.
% Hence, we consider the cost function of the $n$-th CAV as $C(D_n) = pD_n + q$, which relates to the data size of local training.

\textbf{Client-Side Problem.}
On the client side, the objective is to maximize the revenue (i.e., the received reward minus its cost), by optimizing the data size for local training.
Therefore, we formulate the client-side optimization problem of data selection in the $n$-th CAV as follows
\begin{equation}
\begin{split}
\mathbb{P}_0:&\quad \max_{D_n}\quad \alpha_n(1-e^{-\beta_n D_n})-\theta \cdot { C(D_n)}\\
&s.t.\quad  \left\{\begin{array}{lc}
{C(D_n)}/{F_n} \le T_{max},\\
\\
D_{n} \le D_{max},\\
\end{array}\right.
\end{split}
\label{eq:optimization1}
\end{equation}
where we introduce the latency requirement by dividing the computation complexity of local training and the computation capacity of the $n$-th CAV (i.e., $F_n$).
Here, $T_{max}$ is the maximum latency allowance and $D_{max}$ is the maximum data size.
In addition, $\theta$ is the rate per computation capability, e.g., per TFLOPS.
Note that, the reward weight $\alpha$ is given by the server side, when solving this client-side problem.

\textbf{Server-Side Problem.}
On the server side, the objective is to maximize the overall accuracy under the given budget, by optimizing the reward weight of individual CAVs.
Therefore, we formulate the server-side optimization problem of incentive design as follows
\begin{equation}
\begin{split}
\mathbb{P}_1:& \quad \max_{\{\alpha_n, n\in\mathcal{N}\}}\quad \sum\limits_{n\in \mathcal{N}} A(D_n(\alpha_n), \pi_n)  \\
&s.t.\quad  \begin{array}{lc}
\sum\limits_{n\in \mathcal{N}}\alpha_n(1-e^{-\beta_n D_n}) \le M_{max},\\
\end{array}
\end{split}
\label{eq:optimization2}
\end{equation}
where we exclude the $w, N$ in Eq.~\ref{eq:accuracy} without affecting the overall optimization problem, and $M_{max}$ is the total budget.
Here, we consider the non-IID severity $\pi_n$ is known by the $n$-th CAV and fixed in each federated learning round.
Note that, we denote \textbf{sensitivity function} as $D_n(\alpha_n)$, which represents the size of training data in the $n$-th CAV under a given reward weight $\alpha_n$.
This is because, selfish CAVs will determine their contributed size of training data independently, where the underlying calculations are dependent on various factors, e.g., user sensitivity of rewards and computation capacities. 
From the perspective of the server side, the sensitivity function of CAVs is unknown.

\textbf{Challenges.}
The technical challenge of solving the aforementioned problems lies in the unknown sensitivity function of CAVs from the perspective of the server side.
As a result, the server-side problem falls into the black-box optimization domain, where conventional linear, nonlinear, and convex optimization methods cannot work directly.
With the massive number of CAVs in federated learning, it is intractable to employ exhaustive search techniques in practice.
Although Bayesian optimization provides a generic framework to solve blackbox problems, its scalability remains uncertain under thousands of CAVs (if not more) in federated learning.

% The primary challenge in solving this optimization problem lies in the fact that the system is unaware of the sensitivity functions. The system can only iteratively adjust the distributed reward weights based on feedback from the CAVs. This adjustment process aims to incentivize the CAVs to offer the largest possible datasets while ensuring that the rewards remain within a predetermined budgetary constraint. Consequently, this optimization issue is transformed into a black-box optimization problem. Additionally, given that CAVs dynamically participate in the federated learning process, employing exhaustive search techniques to determine the optimal solution is impractical.
% 

% challenges: why the problem is challenging to be solve. 1) unknown sensitivity functions, i.e., blackbox optmization problems; 2)dynamic participant, so exhuastive searching fail

\section{Proposed Solution}

% talk about how to solve the problem in high-level. (and why it may work) (you can refer to abstract, first, second, third steps) 

% you may want to have a block diagram to show the structure of the algo, 
In this section, we introduce the \underline{L}earn-to-inc\underline{e}ntivize \underline{F}ederated learn\underline{i}ng (\emph{LeFi}) algorithm, which allows the server to iteratively learn to allocate reward weights to individual CAVs without prior knowledge. 
Different from vanilla Bayesian optimization, the \emph{LeFi} algorithm works by following three steps.
First, we use a sample-efficient surrogate model to approximate the size of training data under different reward weights for each CAV.
Second, we design a sample-based gradient descent method to update the reward weight of individual CAVs.
Third, we project the reward weights under the total budget.
The server will receive the actual data size of local training in each CAV, which will be used to update the surrogate model continuously.
These steps iterate until the convergence of the algorithm.

% CAVs' sensitivity functions through continuous observation of the interplay between reward weights and dataset sizes. Subsequently, the dataset sizes is predicted, and gradient descent is applied to adjust the reward allocation. The reward weight $\alpha_n$ is iteratively updated, with each update bringing the allocation within the constraints of the system's budget. The operational framework of the \emph{LeFi} algorithm is depicted in Fig \ref{fig:alg_block_diagram}.
% 
% \begin{figure}[!t]
% 	\centering
% 	\includegraphics[width=1.5in]{fig/alg_block_diagram.pdf}
% 	\caption{\small Model accuracy varies in different dataset sizes and Non-IID(Polynomial Surface Fit) }
% 	\label{fig:alg_block_diagram}
% \end{figure}

\subsection{Unknown Function Approximation}
% what is the goal of this subsection, i.e., to learn the function from accumulative observations/feedback
% Due to the system's lack of knowledge regarding the vehicle's sensitivity functions, the system can learn the relationship by observing the size of the dataset provided by the vehicle and rewards previously allocated by the system. Through accumulated observation and learning, an approximate function can be derived that progressively converges towards accurate estimation.

% what are existing methods to solve this small-problem, e.g., maybe linear regression (poor capability), DNN (not sample efficient)

Here, the goal is to find a sample-efficient surrogate model to learn and fit the unknown sensitivity function.
The surrogate model can be updated via accumulative query-and-observation.
In particular, we focus on learning the sensitivity function of individual CAVs, rather than the overall accuracy model.
Although linear and polynomial regression are generally both compute- and sample- efficient, their approximation capabilities are limited to represent complex and unknown functions in selfish CAVs.
In contrast, deep learning (DL) utilizes multiple neurons and multi-layer neural networks to effectively fit nonlinear and high-dim complex functions. 
Nonetheless, DL practically requires a substantial amount of training datasets, with a low sample efficiency.

% so, we use gaussian process regression, then talk about its advantages, 

% put some equation of GPR here, you can refer to our previous papers, but do NOT COPY, you can rephrase
Alternatively, GPR offers several advantages in terms of sample-efficiency and approximation capability. 
Unlike linear regression, GPR does not assume a predefined form of the relationship between inputs and outputs, allowing for the capture of complex, non-linear patterns. 
This non-parametric approach leverages a covariance function, or kernel, which encapsulates assumptions about the function to be learned. 
In addition, GPR can provide a probabilistic prediction along with the uncertainty of the estimation, which is crucial for decision-making processes.

The Gaussian process is defined by its mean function \( m(x) \) and covariance function \( k(x,x') \), which is expressed as:
\begin{equation}
m(x) = \mathbb{E}[f(x)],
\end{equation}
\begin{equation}
k(x,x') = \mathbb{E}[(f(x) - m(x))(f(x') - m(x'))],
\end{equation}
where \( f(x) \) represents the unknown function to learn. 
The choice of the kernel function \( k \) is pivotal as it determines the smoothness and other properties of the functions drawn from the process. 
The resultant predictive distribution for a new input \( x^* \) is then given by a normal distribution characterized by a mean \( \mu(x^*) \) and variance \( \sigma^2(x^*) \), making GPR a robust tool that inherently quantifies the uncertainty of its predictions. 
% This inherent quantification of uncertainty is an asset in real-world applications where predictions are not just point estimates but distributions that reflect real-world variability.

With the GPR, we define its input as the reward weight and the output as the resulting data size per CAV. 
In each iteration, the GPR of CAVs is updated by using all the accumulated reward-to-size pairs in the previous iterations.

\subsection{Reward Allocation Updating}
% what is the goal of this subsection, i.e., with the learned function approximation, we need to solve the problem, 
Here, the goal is to conduct a one-step update on the reward weight of all CAVs given the current GPRs, towards the optimal solution.

First, we deviate the last reward weight $\alpha_n$ with $\triangle$, and predict its mean and std by inferring the GPR as 
\begin{equation}
\mu, \sigma = GPR(\alpha_n + \triangle),
\end{equation}
where $\mu$ and $\sigma$ are the mean and std, respectively.

Second, we estimate the gradient at the last reward weight $\alpha_n$ as
\begin{equation}
\frac{\partial D_n }{\partial \alpha_n} = \frac{ Normal(\mu, \sigma)- GPR(\alpha_n)}{\delta},
\end{equation}
where we sample from the Normal distribution $Normal(\mu, \sigma)$.
Here, we use the sampled value (rather than the mean) to incorporate its variance in the following gradient descent.

Third, we calculate the gradient by following 
\begin{equation}
\triangledown \overline{g}=\frac{\partial A_n }{\partial \alpha_n}=\frac{\partial A_n }{\partial D_n} \cdot \frac{\partial D_n }{\partial \alpha_n},
\label{eq:gradient}
\end{equation}
where,
\begin{equation}
\frac{\partial A_n }{\partial D_n} = 2a D_n + c \pi_n+d,
\end{equation}
where $a$, $c$, and $d$ are constants obtained from polynomial regression of accuracy function in CAVs.

Fourth, we update the last reward weight as follows
\begin{equation}
\alpha_n^{t+1}=\alpha_n^t+\eta \cdot \triangledown \overline{g},
\label{eq:update_alpha}
\end{equation}

Fifth, we project the generated $\alpha_n^{t+1}$ under the total budget (if applicable) as follow
\begin{equation}
    \alpha_n^p =  {M_{max}}/{\sum_{n \in N} \alpha_n(1-e^{-\beta_n D_n})},
\label{eq:scale_factor}
\end{equation}
which happens only if $\sum_{n \in N} \alpha_n(1-e^{-\beta_n D_n}) \ge M_{max}$.

% It is noteworthy that, since the Accuracy model may not be smooth and convex, we utilize not only the mean estimated by GPR to predict the reward weights, but also seek the global optimum by considering both the estimated mean and variance learned from the GPR model, with the intention of accommodating dynamic variations. Specifically, we generate random numbers that follow a normal distribution with the mean and variance provided by GPR as parameters to represent the reward weights.The formula is as follows, where $Z$ is a random number generated from the standard normal distribution:
% % put some equations
% % \begin{equation}
% % f(x)=\frac{1}{\sigma \sqrt{2 \pi}} e^{-\frac{1}{2}\left(\frac{x-\mu}{\sigma}\right)^2}
% % \end{equation}
% \begin{equation}
%     D = \mu + \sigma \cdot Z
% \end{equation}
% % next, project the reward to the budget, with an equation and explanation

\subsection{Selfish Dataset Selection}
% focus on the vehicle side, how to select the dataset
From the perspective of individual CAVs, optimizing the data size of local training is a key approach to maximizing its own revenue.
Selecting a larger data size can generally improve the received reward from the server side, but this also introduces additional costs, which could potentially decrease the revenue. 
Here, we employ the KKT conditions to identify the optimal solution for training data size. 

Initially, we establish the Lagrangian function with the introduction of Lagrange multipliers to account for the imposed constraints.
Let \( \lambda_1 \) be the Lagrange multiplier for the first constraint, and \( \lambda_2, \lambda_3 \) be the multipliers for the inequality constraints \( D_{min} \leq D_n \) and \( D_n \leq D_{max} \), respectively. Moreover, the Lagrange multipliers must fulfill the non-negativity conditions,$\lambda_1, \lambda_2, \lambda_3 \ge 0 $.

The Lagrangian, \( \mathcal{L} \), can be articulated as follows
\begin{equation}
\begin{split}
\mathcal{L}(D_n, \lambda_1, \lambda_2, \lambda_3) &= \alpha_n(1-e^{-\beta_n D_n}) - \theta\cdot{(aD_n+b)} \\
& + \lambda_1 \left(T_{max} - \frac{aD_n+b}{F_n}\right) \\ 
& + \lambda_2(D_n - D_{min})\\
& - \lambda_3(D_n - D_{max}),
\end{split}
\end{equation}
Then, we differentiate \( \mathcal{L} \) with respect to \( D_n \) and set the derivative equal to zero
\begin{equation}
\frac{\partial \mathcal{L}}{\partial D_n} = -\alpha_n \beta_n e^{-\beta_n D_n} - \theta\cdot{a}
 - \lambda_1 \frac{a}{F_n} + \lambda_2 - \lambda_3, \\
% \frac{\partial \mathcal{L}}{\partial D_n} & = 0.
\end{equation}

For the inequality constraints, the following conditions must be satisfied
\begin{equation}
\begin{split}
\lambda_1 \left(T_{max} - \frac{aD_n+b}{F_n}\right) = 0, \\
\lambda_2(D_n - D_{min}) = 0, \\
\lambda_3(D_n - D_{max}) = 0.
\end{split}
\end{equation}

Note that, the data selection problem in individual CAVs is not known by the server side. 
Our proposed algorithm makes no assumption on data selection methods and can be easily extended to accommodate other methods in selfish CAVs.

% along with Tmax, Dmax, Dmin, to determine the data

\begin{algorithm}[!t]
    \caption{The \emph{LeFi} algorithm}\label{alg:LeFi}
    \KwIn{$M_{max},F_n,\theta_n,\pi_n,D_{min},D_{max},\eta$}
    \KwOut{$D_n,\alpha_n$}
    Initialize the reward weight $\alpha_n$ and warm start the GPR model;
    \For{$t = 0,1,...,Rounds$}{
        $/**\;GPR\; Prediction**/$\;
        Build training dataset with $[\alpha, D]$ and train GPR model\;
        Get mean \( \mu(D_n) \) and std \( \sigma \) of deviated dataset size $\alpha +\triangle$ from GPR\;
        Sample dataset size from normal distribution $Normal(\mu, \sigma)$\;
        $/**\;Sample-based\;Gradient\; Descent**/$\;
        Update reward weight $\alpha_n$ based on Eq.~\ref{eq:update_alpha}\;
        $/**\;Project\; Reward\; Weight**/$\;
        \If{$\sum_{n \in N} \alpha_n(1-e^{-\beta_n D_n}) \ge M_{max}$}{
            Project reward weights based on Eq.~\ref{eq:scale_factor}\;
        }
    }
    \Return optimal $D_n$ and $\alpha_n$ for all CAVs\;
\end{algorithm}

\begin{figure*}[!t] % cannot have space
\captionsetup{justification=centering}
  \begin{minipage}[t]{0.32\textwidth}
	\centering
	\includegraphics[width=2.4in]{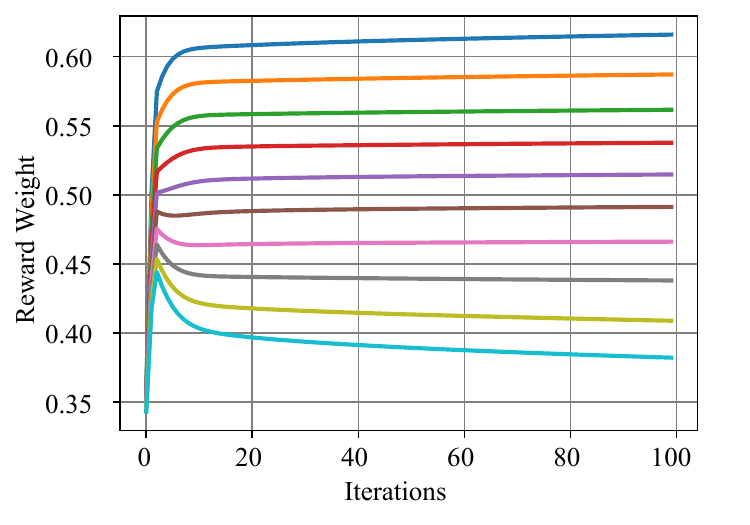}
	\vspace{-0.2in} \caption{\small Convergence of rewards weights.}
	\label{fig:fig_aplha}
  \end{minipage}
  \begin{minipage}[t]{0.32\textwidth}
	\centering
	\includegraphics[width=2.4in]{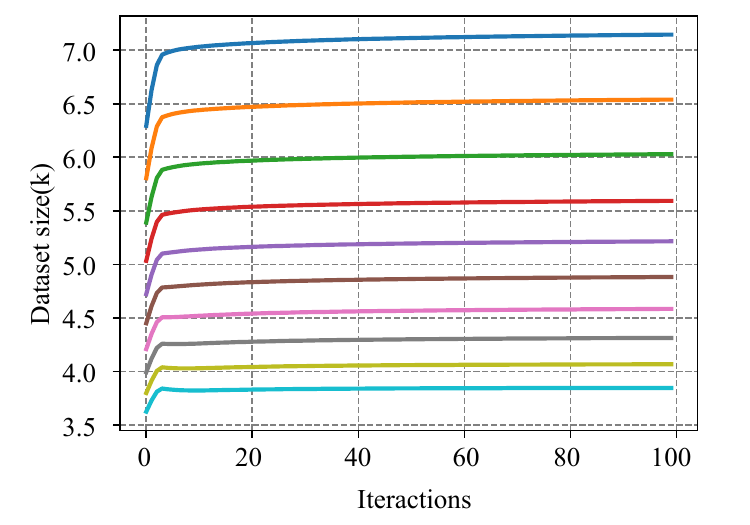}
	\vspace{-0.2in} \caption{\small Convergence of data selection.}
	\label{fig:fig_datasize}
  \end{minipage}
  \begin{minipage}[t]{0.32\textwidth}
	\centering
	\includegraphics[width=2.4in]{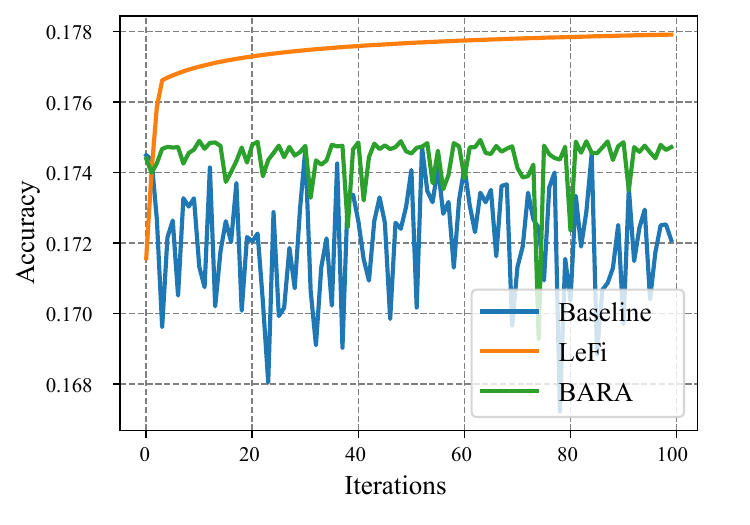}
	\vspace{-0.2in} \caption{\small Convergence of accuracy.}
	\label{fig:fig_a}
  \end{minipage}
\end{figure*}

\subsection{The \emph{LeFi} Algorithm}
Based on the above analysis, we introduce the \underline{L}earn-to-inc\underline{e}ntivize \underline{F}ederated learn\underline{i}ng (\emph{LeFi}) algorithm, whose pseudocode is provided in Alg.~\ref{alg:LeFi}.

The process begins with the initialization of reward weight $\alpha_n$. 
Using the $\alpha$ and the optimal dataset size $D$ obtained from the KKT conditions, we construct the training set and train the GPR of each CAV. 
Next, the GPR model predicts with the last deviated $\alpha_n + \triangle$ and uses the mean and std as parameters to randomly generate $D_n(\alpha_n + \triangle)$ followed by the Normal distribution.
Then, the gradient is calculated and the reward weight of all CAVs is updated accordingly.
If the reward allocated exceeds the given budget, the reward weights are projected back into the range of the budget.
In each iteration, all previously observed values of \(D_n\) and \(\alpha_n\) are used to train the GPR, which would gradually improve its approximation accuracy towards the convergence of the algorithm.

\section{Performance Evaluation}

% \begin{figure*}[!t] % cannot have space
% \captionsetup{justification=centering}
%   \begin{minipage}[t]{0.49\textwidth}
% 	\centering
% 	\includegraphics[width=2.5in]{fig/gap_remaining_gap.pdf}
%     \caption{\small Discrepancy reduction under different user traffic.}
% 	\label{fig:gap_remaining_gap}
%   \end{minipage}
%   \begin{minipage}[t]{0.49\textwidth}
% 	\centering
% 	\includegraphics[width=2.5in]{fig/Discrepancy_reduction.pdf}
%    \caption{\small Discrepancy reduction under different states.}
% 	\label{fig:Discrepancy_reduction}
%   \end{minipage}
% \end{figure*}
% state the parameters used in the simulation. Also justify them via references and sentences.

In this section, we evaluate the proposed algorithm with extensive simulations. 
To obtain the simulation parameters, we profile a CNN-based model under the CIFAR-10 dataset over a desktop with NVIDIA 3080 GPU whose computation capacity is 28.9 FLOPS\cite{nvidia_ampere}.
All the regression models are obtained based on the profiled measurement, including the overall accuracy and the computation complexity.
Based on the pricing of GPU rental services in the market \cite{vast_ai}, we set the cost weight \( \theta \) by generating variations by varying from 0.5 to 2 times 0.9 USD.
Besides, the non-IID severity is uniformly generated between 0.1 and 1.0 for all CAVs. The maximum training time of each round is 200s by default.
The maximum size of training data in each CAV is 10K entries. 
We conduct 2-order polynomial regression to obtain the accuracy model $A(D_n,\pi_n) = aD_n^2+b \pi_n^2 + cD_n\pi_n +dD_n + e\pi_n+f$ with root mean square error (RMSE) 0.03, whose parameters are $a =-0.000152,b=0.071,c=-0.00117,d=0.0151,e=0.011,f=0.073$.
The default number of CAVs is 10, the budget is 5 USD, and the latency requirement is 1 minute. 

% The step size of gradient descent is 1.0.

We compared with the following algorithms\footnote{We omit DRL-based solutions because they can hardly learn in very limited Server-to-CAV interactions (i.e., maximum 100).}: 
\begin{itemize}
    \item \textbf{Baseline}: the \emph{Baseline} method uses random search to find the combination of reward weight of all CAVs, which serves as the baseline performance.
    \item \textbf{BARA}: the \emph{BARA} method~\cite{yang2023bara} uses Bayesian optimization to optimize the reward weight of all CAVs, where a GPR is created to approximate the correlation between all the weights of CAVs and the overall accuracy of the global model. The expected improvement (EI) is used for the acquisition function.
\end{itemize}

\begin{figure*}[!t] % cannot have space
\captionsetup{justification=centering}
  \begin{minipage}[t]{0.32\textwidth}
	\centering
	\includegraphics[width=2.4in]{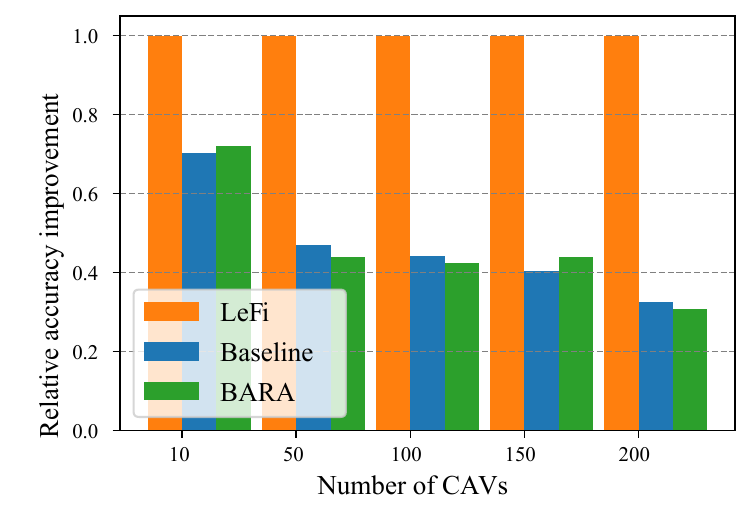}
	\vspace{-0.2in} \caption{\small Relative accuracy improvement under different number of CAVs.}
	\label{fig:fig_scale_scale}
  \end{minipage}
  \begin{minipage}[t]{0.32\textwidth}
	\centering
	\includegraphics[width=2.4in]{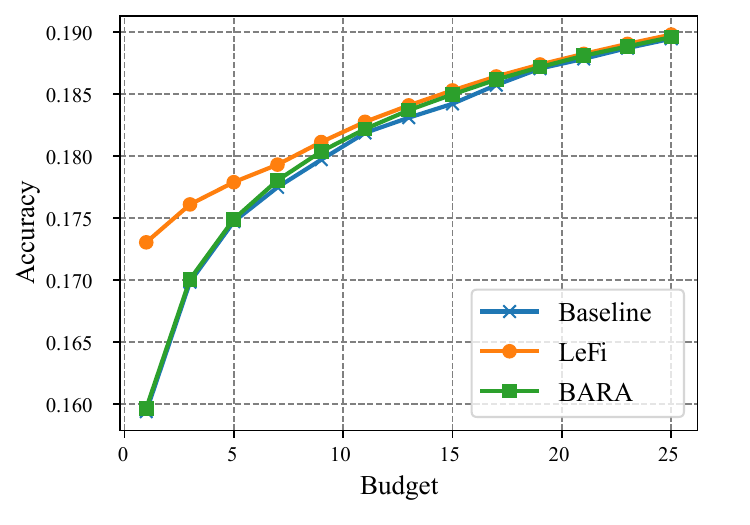}
	\vspace{-0.2in} \caption{\small Accuracy under different budget.}
	\label{fig:fig_budget}
  \end{minipage}
  \begin{minipage}[t]{0.32\textwidth}
	\centering
	\includegraphics[width=2.4in]{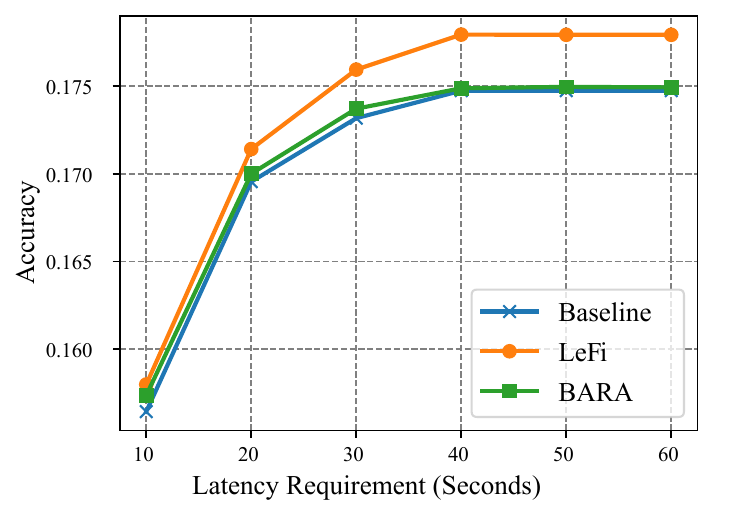}
	\vspace{-0.2in} \caption{\small Accuracy under different latency requirement.}
	\label{fig:fig_tmax}
  \end{minipage}
\end{figure*}

\textbf{Convergence.}
We show the convergence of the \emph{LeFi} algorithm, in terms of reward weights and data sizes in Fig.~\ref{fig:fig_aplha} and Fig.~\ref{fig:fig_datasize}, respectively.
It can be seen that the reward weights (server-side) and data sizes (client-side) converge in about only 20 iterations.
Notably, the reward weight of all CAVs increases in the very beginning (e.g., 5 iterations) and then diverges accordingly, which is because the total budget is reached, and the projection is invoked to ensure the budget for all CAVs.
As the \emph{LeFi} algorithm converges, we can see that the optimal reward weights are different for all CAVs, which is mainly attributed to their heterogeneous sensitivity functions. 
In addition, we show the convergence of overall accuracy under all the solutions in Fig.~\ref{fig:fig_a}.
As compared to \emph{Baseline}, the \emph{BARA} can generally achieve better accuracy performance, because it re-uses the previous observations for future searching.
As compared to other solutions, the \emph{LeFi} algorithm shows a very smooth convergence and eventually higher accuracy performance throughout the iterations.
Note that, although the absolute numeric difference is relatively small, such improvement in the accuracy of the global model is non-trivial.
These results verify the superior convergence of the \emph{LeFi} algorithm.

\textbf{Scalability.}
We show the scalability of the \emph{LeFi} algorithm in Fig.~\ref{fig:fig_scale_scale} by evaluating its accuracy performance under different numbers of CAVs.
Here, the metric of relative accuracy improvement (RAI) is defined as the ratio between the accuracy improvement of a solution and the \emph{LeFi} algorithm.
The accuracy improvement denotes the achieved accuracy increase of the \emph{LeFi} algorithm as compared to arbitrary reward weights.
In general, a higher RAI indicates that the solution can achieve better accuracy. 
As we can see, the RAI of both \emph{Baseline} and \emph{BARA} are below 1.0, which means they are less comparative with the \emph{LeFi} algorithm.
In addition, their RAI difference becomes larger, as the increment of the number of CAVs, which suggests they are less scalable as compared to \emph{LeFi} algorithm.
When there are a massive number of CAVs in federated learning, the searching space for both \emph{Baseline} and \emph{BARA} increases exponentially, which decreases their effectiveness in finding the optimal reward weights of all CAVs.

\textbf{Adaptability.}
We evaluate the adaptability of the \emph{LeFi} algorithm under varying budget and latency requirements in Fig.~\ref{fig:fig_budget} and Fig.~\ref{fig:fig_tmax}, respectively.
As more budget allows higher data sizes in all CAVs, the accuracy performance of all solutions is increased accordingly, as shown in Fig.~\ref{fig:fig_budget}.
When the budget is no longer a significant constraint, all CAVs would like to contribute all of their training data, which is the main reason why all solutions share similar accuracy when the budget is 25 USD.
However, we see that the \emph{LeFi} algorithm can achieve higher improvement than other solutions, especially when the budget is smaller (e.g., 1).
As a more strict latency requirement is enforced, we see the accuracy performance of all solutions generally reduce in Fig.~\ref{fig:fig_tmax}, because CAVs cannot contribute more training data to match the allocated reward weights.
In the meantime, the \emph{LeFi} algorithm typically achieves higher accuracy gain than other solutions, especially when the latency requirement is loosened.
These results justify the better adaptability of the \emph{LeFi} algorithm under different conditions in federated learning.

\section{Related Work}
% To address this, Chai \emph{et al.}~\cite{chai2020hierarchical} employ game theory and blockchain to encourage secure knowledge sharing by designing a transactional market-based approach. Similarly, issues of resource allocation are tackled by Nazih \emph{et al.}~\cite{nazih2020incentive} through a game-theoretic model optimizing vehicular fog nodes' interactions and computing resource sharing. Montazeri \emph{et al.}~\cite{montazeri2022distributed} integrate FL to bolster the data processing capabilities and security of Road Side Units (RSUs), applying contract theory to develop a profitable incentive mechanism. In the Internet of CAVs (IoV), Saputra \emph{et al.}~\cite{saputra2021dynamic} propose employing Smart CAVs (SVs) as federal members to enhance network functionality with minimal data exposure. They devise a contract-based policy focusing on the optimal selection of SVs and payment contracts, contingent on the data quality, which, notably, lacks a mathematical definition in their study. 
% Additionally, Pokhrel \emph{et al.}~\cite{pokhrel2020federated} suggest an autonomous, blockchain-powered FL model that incentivizes data contribution through a reward system calibrated to the revenue of the data sample size.

The incentive mechanism in federated learning has been extensively investigated in recent years.
Chai \emph{et al.}~\cite{chai2020hierarchical} employed game theory and blockchain to encourage secure knowledge sharing by designing a transactional market-based approach.
Saputra \emph{et al.}~\cite{saputra2021dynamic} proposed employing Smart CAVs (SVs) as federal members to enhance network functionality with minimal data exposure. They devise a contract-based policy focusing on the optimal selection of SVs and payment contracts, dependent on the data quality.
Yang \emph{et al.}~\cite{yang2023bara} introduced BARA, a Bayesian optimization-based algorithm for dynamic reward budget allocation in federated learning, aimed at optimizing model utility and accuracy. This novel approach effectively handles the complex relationship between reward allocation and model performance in federated learning environments.
Weng \emph{et al.}~\cite{weng2021fedserving} presented FedServing, a federated prediction serving framework that incentivizes truthful machine learning predictions through a Bayesian game theory-based mechanism and enhances accuracy using truth discovery algorithms. This framework also ensures model privacy and exchange fairness by employing blockchain and Trusted Execution Environments (TEEs). 
Toyoda \emph{et al.}~\cite{toyoda2020blockchain} introduced a mathematical framework using contest theory to analyze and optimize reward distribution in a model updating environment. Emphasizing Bayesian-Nash incentive compatibility ensures that workers maximize their payoffs with no better alternative strategies, and it establishes an optimal reward distribution policy. 
% The analysis treats each reward distribution round uniformly, simplifying the theoretical exploration of worker contribution and payoff maximization within this Bayesian framework.
However, existing works either assume the known sensitivity function of all CAVs or cannot scale to the massive number of CAVs (thousands if not more).
In this work, we proposed the high-scalable \emph{LeFi} algorithm to learn and incentivize heterogeneous CAVs without the need for prior knowledge of their sensitivity functions.

\section{Conclusion}
In this work, we proposed a new \emph{LeFi} algorithm to incentivize the participation of heterogeneous CAVs in federated learning.
The key observation is that the sensitivity function of CAVs is generally unknown by the server side, due to a variety of practical factors, such as computation capacity and monetary sensitivity.
Our key idea to solve this challenge is to develop a sample-efficient surrogate model to learn and approximate the unknown sensitivity function, from accumulated Server-to-CAV interactions.
Then, we design a sample-based gradient descent method to iteratively update the reward allocations and apply projection if needed to meet the budget.
The simulation results show that our proposed algorithm can significantly outperform existing solutions, in terms of accuracy, scalability, and adaptability.

% \section*{Acknowledgement}

\bibliographystyle{IEEEtran}
\bibliography{ref/qiang, ref/ref}

\end{document}